\documentclass[letter]{aa} 
\usepackage{graphicx}
\usepackage{natbib}
\usepackage{ulem}
\usepackage{txfonts}
\begin{document}
   \title{Low-lying magnetic loops in the solar internetwork}


   \author{M. J. Mart\' inez Gonz\'alez \inst{1,2}, M. Collados \inst{2}, B.
Ruiz Cobo \inst{2} \and S. K. Solanki \inst{3}}

   \authorrunning{M. J. Mart\' inez Gonz\'alez et al.}

   \institute{LERMA, Observatoire de Paris-Meudon, 5 place Jules Janssen, 92195
Meudon, France\\
   \email{Marian.Martinez@obspm.fr}
         \and
	      Instituto de Astrof\' isica de Canarias,
              V\' ia L\'actea S/N, 31200 La Laguna, Spain\\
              \email{mcv@iac.es, brc@iac.es}
         \and
             Max-Planck-Institut f\"ur Sonnensystemforschung,
              Max-Planck-Str. 2, 37191 Katlenburg-Lindau, Germany\\
             \email{solanki@mps.mpg.de}}

   \offprints{Marian.Martinez@obspm.fr}

   \date{}

 
  \abstract
   {}
   {The aim of this work is to study the structure of the magnetic field vector
in the internetwork and search for the presence of small-scale loops.}
   {We invert 1.56 $\mu$m spectropolarimetric observations of internetwork
regions at disc centre by applying the SIR code. This allows us to recover the atmospheric 
parameters that play a role in the formation of these spectral lines. We are mainly interested 
in the structure of the magnetic field vector.}
   {We find that many opposite polarity elements of the internetwork are
connected by short ($2-6''$), 
low-lying (photospheric) loops. These loops connect at least the $10-20$ \% of
the internetwork flux visible in 
our data. Also we have some evidence that points towards a dynamic scenario
which can be produced by the emergence of internetwork magnetic flux.}
   {}

   \keywords{Sun: magnetic fields --- Sun: atmosphere --- Polarization --- Methods: observational}

   \maketitle
%

\section{Introduction}

Magnetic fields appear on the solar surface in the form of bipolar regions, with
the two polarities connected by magnetic field lines in the shape of loops. If
the loops are sufficiently large they are traced in H$\alpha$ (arch filament
systems) or EUV or X-ray images (transition region and coronal loops). 
There is also mounting evidence for the presence of myriads of small loop-like
structures in the upper quiet atmosphere \citep[e. g.][]{feldman_99} and in the
photosphere \citep{lites_96, pontieu_02}, and the presence of such loops has
been invoked by the complex mixture of magnetic polarities typical of the quiet
Sun. Direct measurements of the full magnetic structure of a loop are extremely
rare, being basically restricted to chromospheric loops in emerging flux regions
\citep{sami_03}. Here we trace the field of low-lying loops and present evidence 
that they are common in the internetwork quiet Sun.

Our knowledge of the magnetic field in the internetwork has evolved rapidly in
recent years. The magnetic features present in internetwork regions, as seen by
the Zeeman effect, are unresolved, occupy a very small portion of the
resolution element ($1-2$ \%) and have field strengths of a few hundred G
\citep{keller_94, lin_95, lin_99, khomenko_03, arturo_06, marian_spw4,
andres_07}. Probably the rest of the resolution element is filled by magnetic fields that currently are only detectable by means of the Hanle effect \citep{stenflo_87} or the 
Zeeman sensitive Mn\,{\sc i} line at $1.52$ $\mu$m \citep{andres_07}. If we assume that such 
a hidden magnetic field is tangled at subresolution scales and that 
it occupies the whole resolution element, the most sophisticated Hanle effect determinations 
measure a mean magnetic field of 100 G \citep{javier_04} while previous estimates give a 
much weaker field \citep{faurobert_01}. In this type of studies, the spatial
resolution is given by the slit length, since the signal is averaged over the
whole slit in order to obtain an adequate signal to noise ratio. In addition,
some works find empirical evidence of this tangled nature \citep{stenflo_87, rafa_04, andres_07}. 
The present paper extends the knowledge of the quiet Sun by revealing
the connectivity of magnetic features in the internetwork and partly its three
dimensional structure.

\section{Observations and data reduction}

\begin{figure}[!t]
\centering{
\includegraphics[width=3.6cm]{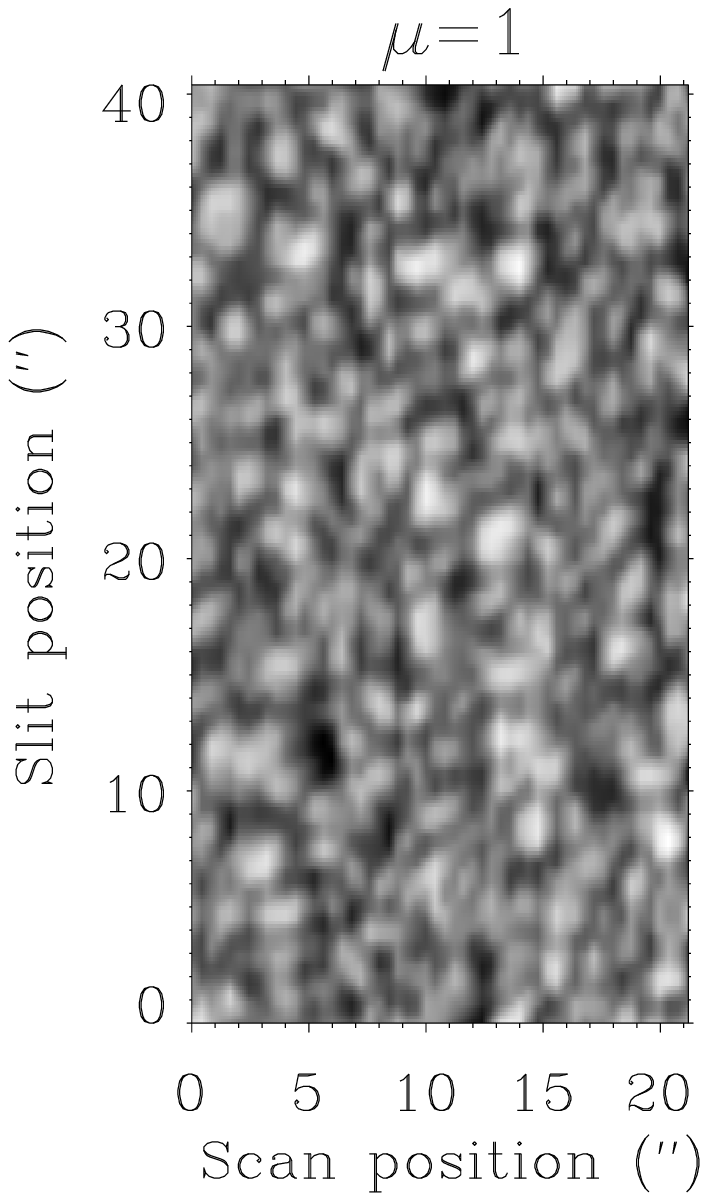}
\hspace{1.6cm}
\includegraphics[width=3.6cm]{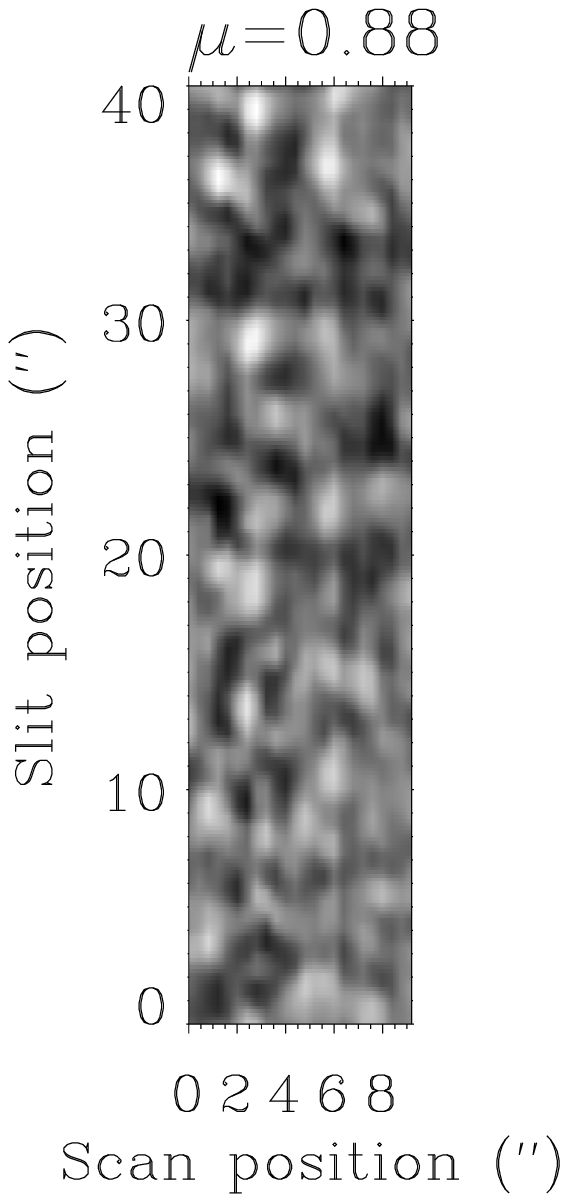}}
\caption{Continuum maps of the observations taken at disc centre (left) and at
$\mu=0.88$ (right).} 
\label{mapas_obs} 
\end{figure}

The observational data consist of two scans of quiet Sun regions. One of them
was located at disc centre and the other one at $\mu=0.88$. 
They were observed on the 29th and the 30th
of July 2000, respectively. All four Stokes parameters of the pair of Fe\,{\sc
i} lines at 1.56 $\mu$m were recorded using the Tenerife Infrared Polarimeter
\citep{manolo99}  installed at the Vacuum Tower Telescope at El Teide
observatory. In both cases, the integration time at each slit position was
\hbox{$1$ min} which allowed us to achieve a noise level in the polarization
profiles of \hbox{$4.1\times 10^{-4}$ I$_\mathrm{c}$} in the observation at disc
centre and \hbox{$2.5\times 10^{-4}$ I$_\mathrm{c}$} in the one at $\mu=0.88$.
Here, I$_\mathrm{c}$ is the continuum intensity. 
In Fig. \ref{mapas_obs}, the I$_\mathrm{c}$ maps are shown for both scanned
areas. The slit was oriented vertically and scanned horizontally. We used a
correlation tracker \citep{ballesteros_96,schmidt_95} to correct for image motions due to
the seeing. The continuum intensity contrast of both data sets is 1.4
\% and 2.0 \%, respectively, for the maps at $\mu=0.88$ and $\mu=1$. In both maps,
the granulation pattern is clear and the spatial resolution was estimated to be
of the order of $1''$. This was computed using the Fourier transform of the
I$_\mathrm{c}$ images and choosing the frequency where the power was 10 times
above noise. In both cases the pixel size, the slit width and the scan step were
all $0.4''$. 

The standard reduction of the data included dark current and flatfield
corrections. The polarimetric calibration was performed with the aid of
polarization optics of known properties, located at an appropriate place in the
optical path. Any residual crosstalk between the Stokes parameters, because of
the uncalibrated coelostat mirrors, were removed using  statistical techniques
described in \citet{manolo03}. After that, a statistical procedure called
Principal Components Analysis (PCA) was applied to reduce the noise level of the
polarization profiles. This consists in creating a basis with the eigenvectors of the 
correlation matrix of the observations in which they can be represented. Then we can eliminate from the basis those eigenprofiles that have no information about the real 
signals and are dominated by noise and reconstruct the observations with this shortened base. This is feasible because of the fast decrease of the eigenvalues and allows us to increase the noise level of the observations.

After denoising with PCA, the noise level of Stokes V is \hbox{$6.5\times
10^{-5}$ I$_\mathrm{c}$} and $9.0\times 10^{-5}$ I$_\mathrm{c}$ for the
observations at $\mu=1$ and $\mu=0.88$, respectively. Stokes Q has a noise level
of \hbox{$1.1\times 10^{-4}$ I$_\mathrm{c}$} and \hbox{$8.5\times 10^{-5}$
I$_\mathrm{c}$} and Stokes U $2.0\times 10^{-4}$ I$_\mathrm{c}$ and $5.0 \times
10^{-5}$ I$_\mathrm{c}$ at disc centre and at
$\mu=0.88$, respectively.

\section{Analysis of the data}

We obtain the magnetic field vector by inverting the four Stokes profiles using
the SIR\footnote{Stokes Inversion based on Response functions} code
\citep{basilio_92}. We perform a two component inversion based on a magnetic
atmosphere that occupies a fraction of the resolution element and a field free
one filling the rest of the space.  Including
two components in the inversion increases the number of free parameters. Then,
in order to end up with an adequate number of degrees of freedom
\citep[see][]{andres_hector_07} we chose the majority of the variables to be
constant with height. The microturbulent velocity in
both components, the line of sight flow velocity of the magnetic component, the
magnetic field strength and the azimuth of the magnetic field vector are assumed
to be constant with height. The macroturbulent velocity is forced to be the same
in both atmospheres. The macroscopic velocity of the non magnetic atmosphere has
3 nodes and the inclination of the magnetic field vector has 2 nodes. The temperature stratification of each atmosphere was allowed to vary independently with $5$ nodes. 

\begin{figure*}
\center
\includegraphics[width=5.18cm]{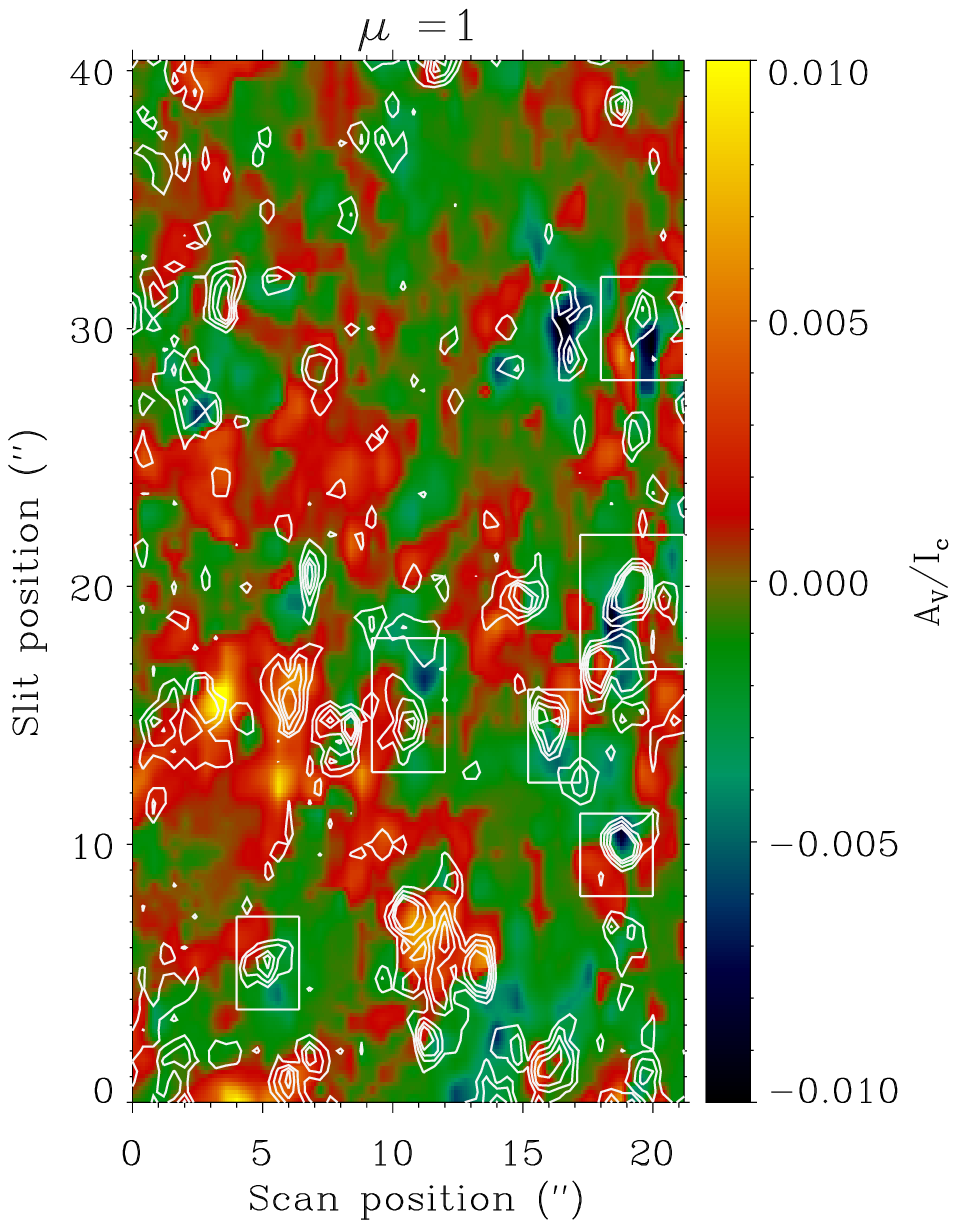}
\hspace{5cm}
\includegraphics[width=5.18cm]{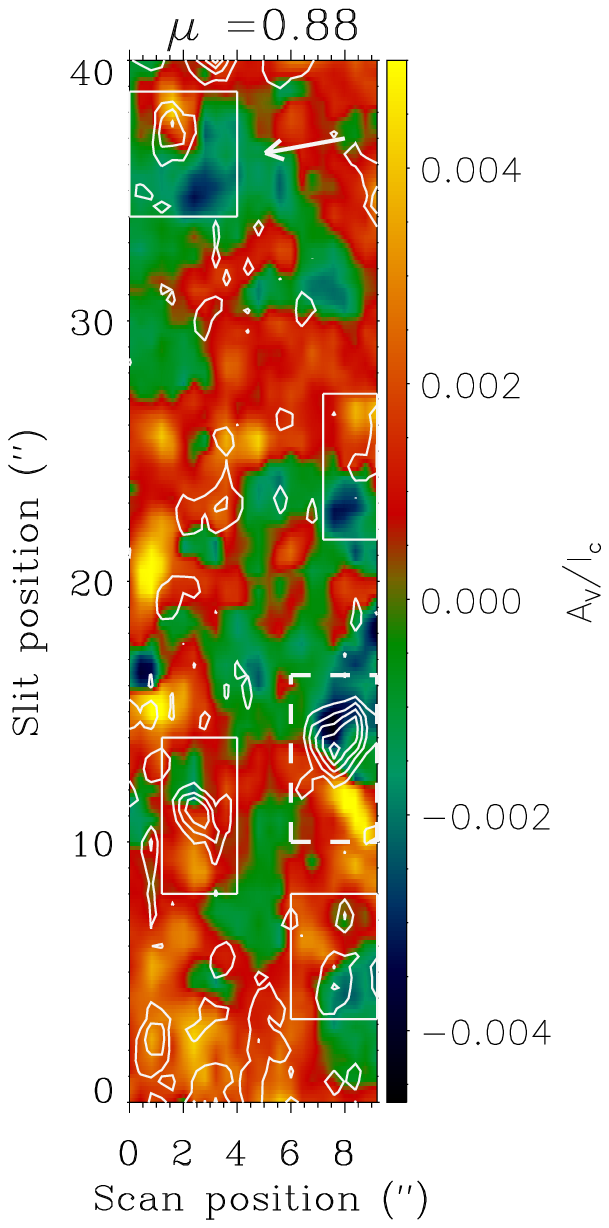}
\caption{Maps of the amplitude A$_\mathrm{V}$ of circular polarization (colour scale) of the observations
taken at disc centre (left) and $\mu=0.88$ (right). The white contours represent
the degree of linear polarization $\sqrt{A^2_Q+A^2_U}$ having the following
values: $1.5\times 10^{-3}, 2\times 10^{-3}, 2.5\times 10^{-3}, 3\times 10^{-3}$
I$_\mathrm{c}$ for the map at disc centre and additionally at $1\times 10^{-3}$
for the map at $\mu=0.88$. The white arrow in the $\mu=0.88$ map points towards the position of the disc centre.}
\label{mapas_pol}
\end{figure*}

In Fig. \ref{mapas_pol} we show the maps of the amplitude of circular
polarization for both scans. Superimposed to them, some contours of equal
amplitude of linear polarization ($\sqrt{A_Q^2+A_U^2}$) are plotted ($A_Q$ and
$A_U$ represent the maximum absolute amplitude of the Stokes $Q$ and $U$
profiles). In both maps there are numerous examples of a linear polarization
signal lying in between two opposite polarity circular polarization signals.
Small low-lying magnetic loops are expected to show just such a signature: the
magnetic field vector points up on one side of the structure, is horizontal in
the middle and points down on the other side. We found $6$ clear cases of such
structures in the map at disc centre and $5$ in the map at $\mu=0.88$. They are
enclosed by the white rectangles in Fig. \ref{mapas_pol}. In the examples in the $\mu=0.88$ map all the linear polarization is slightly displaced towards one of the circular polarization signals. In the disc centre map this is not a systematic behaviour. It might be a projection effect or the foot points having different field strengths.

We discuss now in detail the results from inversions of the clearest loop identified by the thick
dashed-line rectangle in the observation at $\mu=0.88$ (the others show similar properties and geometries). The inclination and azimuth of the magnetic field vector recovered from the inversions in the line-of-sight reference frame are transformed to the local reference frame (without solving the $180^\mathrm{o}$ degrees azimuth ambiguity).

\begin{figure*}
\center
\includegraphics[width=5.3cm]{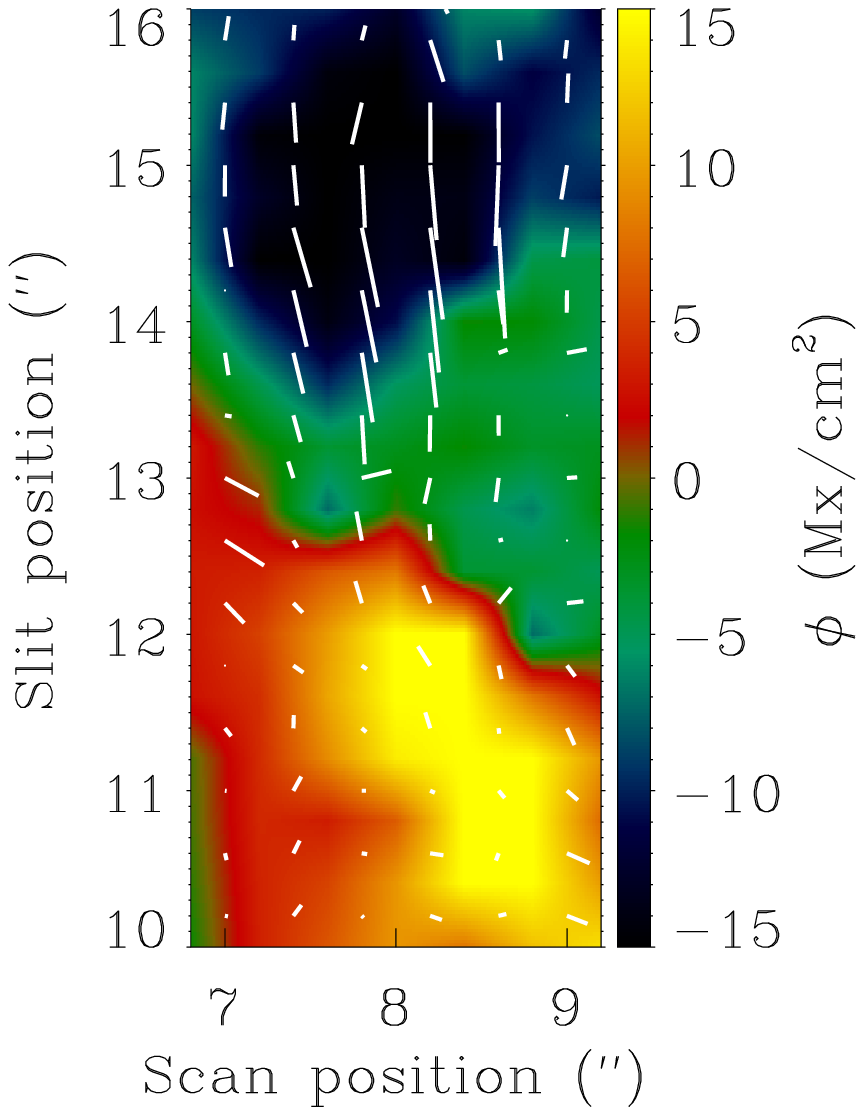}
\hspace{0.7cm}
\includegraphics[width=5.3cm]{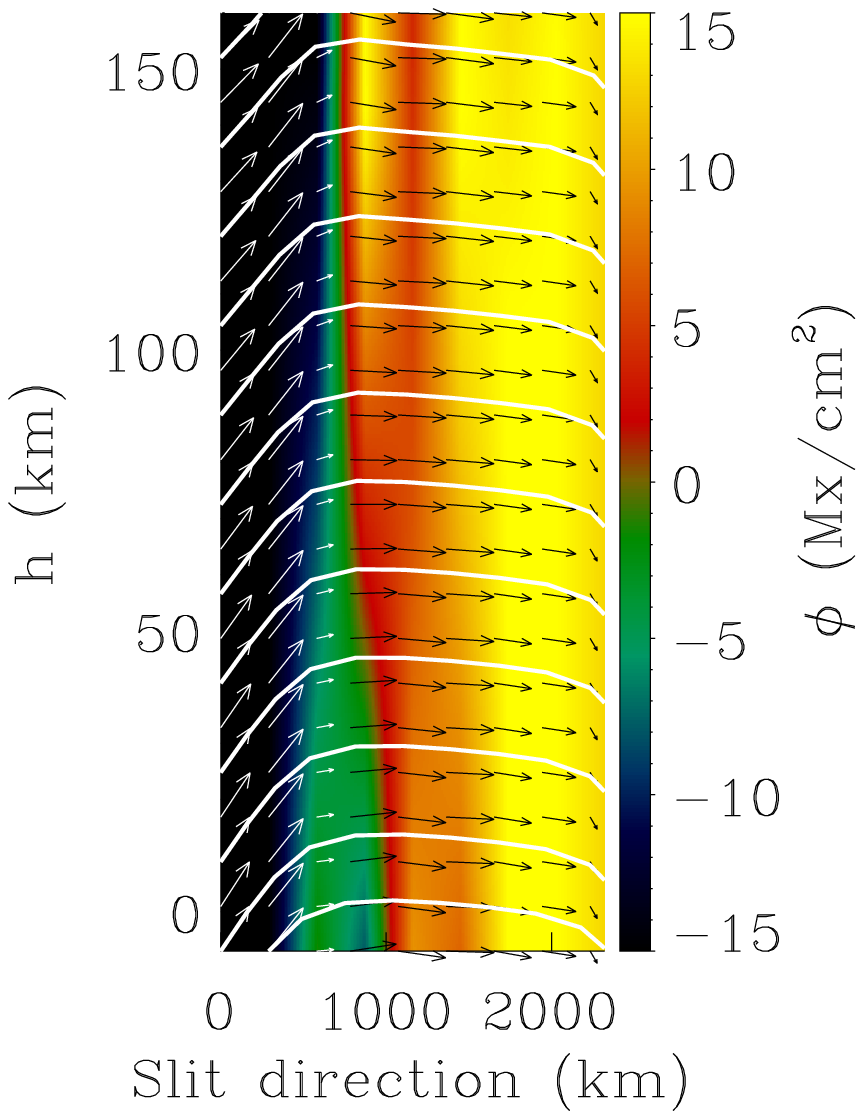}
\hspace{0.4cm}
\includegraphics[width=5.3cm]{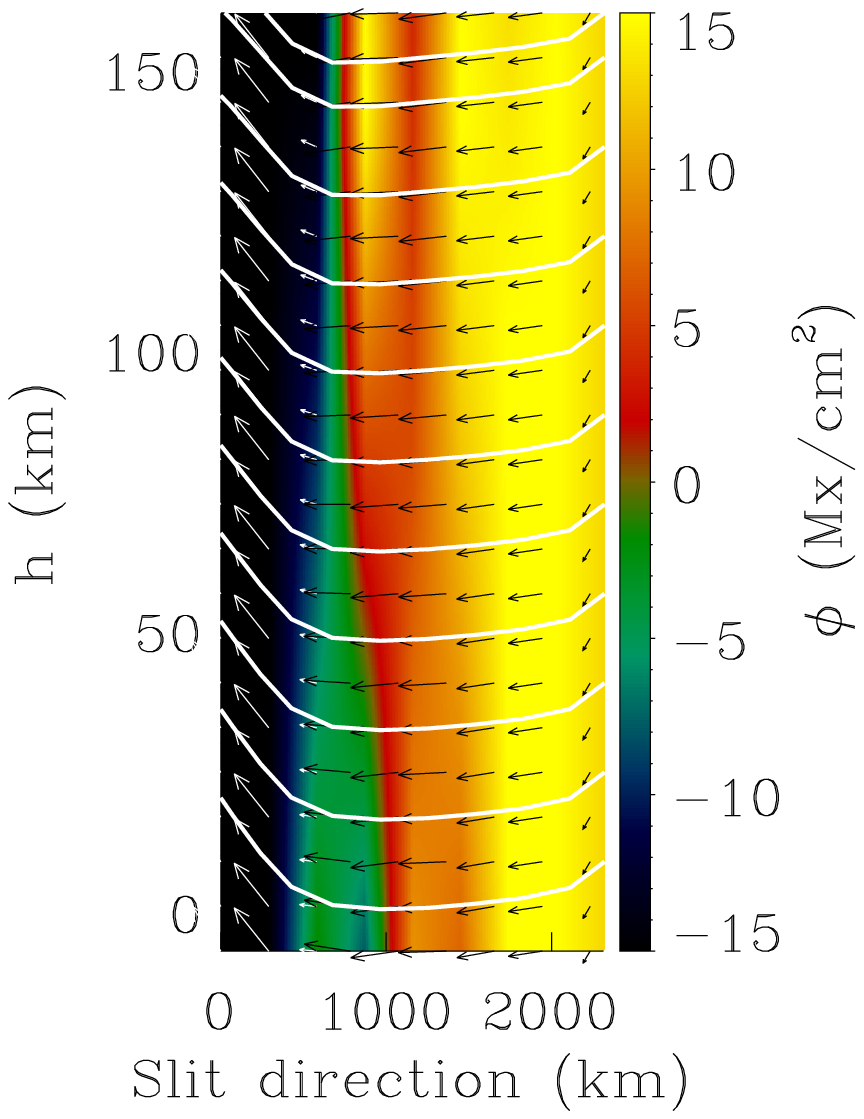}
\caption{Left panel: Map of the vertical magnetic flux density of the loop selected as an
example (marked with a dashed square in the \hbox{$\mu=0.88$} circular
polarization map). The white lines represent the horizontal component of the magnetic flux
vector. Centre and right panels: magnetic field lines computed for the two
allowed magnetic field configurations,
that differ by $180^\mathrm{o}$ in azimuth, giving rise to an $\Omega$- and a U-loop
structure, respectively. The inclination and azimuth of the magnetic field vector have been translated to the local reference system.}
\label{recons_loop}
\end{figure*}

In the left panel of Fig. \ref{recons_loop}, the horizontal component of the magnetic flux 
vector is presented together with the vertical magnetic
flux density for the thick dashed rectangle in Fig. \ref{mapas_pol}. 
The white lines 
in the left panel of Fig. \ref{recons_loop} represent the direction of the 
magnetic field vector. 
The values of the background flux density clearly correspond to  internetwork regions in all the pixels
containing the loop-like structure. The horizontal component of the magnetic
field vector in the pixels in between the two opposite polarities is
approximately directed along the line that joins them. In the central and right
panels of Fig. \ref{recons_loop}, we plot the magnetic field lines on a vertical
cut through the atmosphere along the line joining the two polarities. The
vertical magnetic flux density is indicated by the colour scale and the superimposed
arrows represent the magnetic field vectors. The azimuth values differ by
$180^\mathrm{o}$ in the two panels, both solutions being consistent with the
observed spectral profiles. Note that the inclination of the magnetic field
vector is nearly height independent even if we allow it to vary. The magnetic field lines have been computed as those lines that are parallel to the vector field. In each slit position in centre and right panels the geometrical heights have been displaced to have the same gas pressure at h=0. The reference gas pressure value is the mean value at h=0. The applied displacements are small, having a mean value of $20.5$ km. The
$\Omega$-loop (centre) or U-loop (right) structure is evident: the magnetic
field is almost vertical at the points where the strongest Stokes V signals are
(at the loop foot points) and is horizontal in the transition between one
polarity and the other. Recently \citet{arturo_guillaume_06} have found
U-loops in the photosphere below a filament structure by 
solving the $180^\mathrm{o}$ ambiguity inherent to the Zeeman effect, suggesting that these 
structures are at least allowed by the physics that
  drives the solar atmosphere. In our case, unfortunately, we do not 
have any observational constraint to choose between the two different 
solutions showed in Fig. \ref{recons_loop}.

There are many bipolar regions potentially harboring low-lying loops in the maps
in Fig. \ref{mapas_pol} but only those boxed in are as clear as the example
shown in Fig. \ref{recons_loop}. This means that in all these cases the
horizontal component of the magnetic vector is mainly directed along the line
joining the opposite magnetic  polarities and that the inclination is almost
vertical in the foot points of the structure and rather horizontal in between.
The observed loops are rather flat, with a foot point separation of 
2300 km and with at least some of the field lines at the loop top lying in the
height range of formation of the 1.56 $\mu$m lines (i. e. only a few hundred km
above the foot points). All cases also share some other properties. The
magnetic field strength is weak, being below $500$ G in the majority of the
cases. There are only very few points reaching higher values as $800-1000$ G.

\begin{figure}
\hspace{-0.8cm}
\includegraphics[width=10cm]{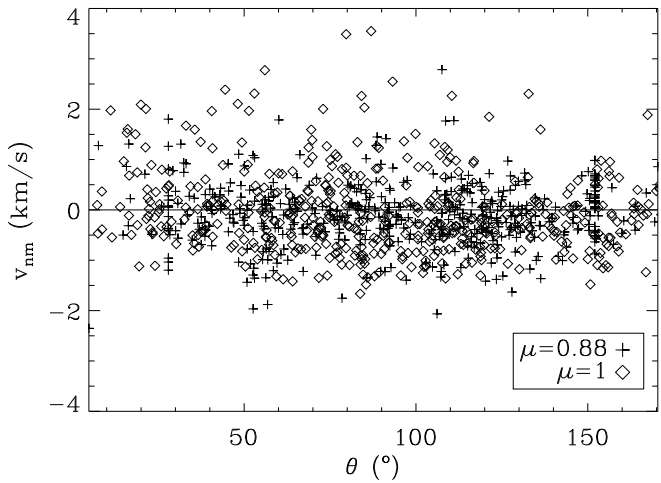}
\caption{Line-of-sight bulk velocity of the field free component plotted versus the inclination of the magnetic field inclination with respect to the local vertical. Positive values correspond to downflows and negatives to upflows.}
\label{vel_th}
\end{figure}

The line of sight plasma flows in the magnetic atmosphere are correlated with
the continuum intensity: bright structures in the continuum image are associated
with upflows in the magnetic component, dark structures with downflows. The foot
points of the magnetic structures are located equally probably in upflows or in
downflows as seen in Fig. \ref{vel_th}. The more vertical fields are located at the footpoints of the structure and we can see that they are found either in upflows or downflows. Also the more horizontal fields are lying both in upflows and downflows. This could be indicative of a dynamic scenario. If the time
scale of the evolution of such structures is larger than the convective turnover time of
granulation, one would expect $\Omega$-loops to be anchored in intergranular lanes
(where downflows dominate) and to be lying over granules. For U-loops the situation is 
expected to be reversed, however. Consequently, we may be seeing a mixture of $\Omega$ and 
U-loops.

As seen in Fig. \ref{recons_loop}, the magnetic flux density of the  selected
loop has typical values of the internetwork. This is also 
true for the rest of the loops boxed in Fig. \ref{mapas_pol}. The magnetic flux 
density in the pixels occupied by loops follow the same
distribution as the pixels in the rest of the internetwork area. 
Also the magnetic field strength distribution is the same as that of the 
rest of the area covered by the typical internetwork. The fact that both the selected 
loops and the rest of the observed area share the same properties suggest that the 
internetwork could harbour many of these small low-lying loops, most of 
them remaining undetectable in our observations.

\section{Discussion and conclusions}

We have detected $11$ examples of low-lying (i. e. photospheric) loops in two
scans of internetwork regions in the quiet Sun. These loops may be related to
the horizontal features observed by \citet{lites_96} and also the emerging flux
in form of small bipoles observed by \citet{pontieu_02} using the visible lines
at $630$ nm. In our case, we have more sensitive spectropolarimetric
observations (very magnetically sensitive infrared lines 
and very low noise in both linear and circular polarization) which allow us to
reconstruct the topology of the magnetic structures. Although we could not
follow the evolution of such loop-like structures, one may speculate that the  physical processes that give rise to these loops are small scale magnetic flux emergence in the internetwork quiet Sun. 

Concerning the magnetic flux carried by the loop-like structures that we
{have identified}, we find that at least $6.1$\% (disc centre) and $25.8$\%
($\mu=0.88$) of the magnetic flux in the whole map is emerging in the form
of small scale magnetic loops. On average, this means that at least
\hbox{10-20\%} of the magnetic flux in the solar internetwork is in the form of
low-lying loops at any given time. This is a lower limit since further such
features may have escaped detection due to the weakness of the associated linear
polarization signals. One implication of the fact that these loops are so
low-lying is that the magnetic flux they are connecting is unlikely to reach the
chromosphere and higher layers. Unless we are observing a transient phase this
implies that possibly much of the magnetic flux does not rise above the
photosphere. This point deserves further attention in view of the suggestions of
\citet{javier_04} that a tangled field could provide the clue to understanding how the 
solar chromosphere and corona are heated.

\begin{acknowledgements}
This research has been funded by the Spanish Ministerio de Educaci\'on y
Ciencia 
through project AYA2004-05792. Part of this work was done when M. J. Mart\' inez
Gonz\'alez was visiting the Max Planck Institute in Lindau and she would like to
express her gratitude to the staff for their warm hospitality. The authors acknowledge the comments of the anonymous referee which helped to improve the letter and strengthen the conclusions. 
\end{acknowledgements}


\end{document}